\documentstyle[epsf]{ichep}
\begin{document}
\title{Nambu mass hierarchies in low energy \\string models}
\author{Emilian Dudas}
\address{Laboratoire de Physique Th\'eorique et Hautes Energies
\\Universit\'e Paris-Sud, Bat. 211, F-91405 Orsay Cedex, France}
\abstract{This paper explores a recent idea of Nambu to generate
hierarchies among Yukawa couplings in the context of effective
supergravity and superstrings models. The Yukawa couplings are
homogeneous functions of the moduli and a geometrical constraint
between them with a
crucial role in the Nambu mechanism is found in a class of models of
no-scale type. The Yukawas are dynamical variables at low energy to be
determined by a minimization process.}
\twocolumn[\maketitle]
%%%%%%%%%%%%%%%%%%%%%%%%%%%%%%%%%%%%%%%%%%%%%%%%%%
\section{Nambu mass hierarchies}
A mistery of the Standard Model is the difference between the mass of
the top quark and the mass of the other fermions. The top quark mass is
roughly of the order the electroweak scale $v \simeq 250 GeV$, whereas
in a first approximation all the other fermions are massless. No
definite solution of this puzzle was found and so new ideas are
necessary.

An interesting idea was recently proposd by Nambu \cite{nambu}. The
Yukawa couplings are regarded as dynamical variables to be determined
by minimizing the vacuum energy density. All the other parameters are
held fixed, including the vev's of the scalar fields. If $\Lambda$ is
the cut-off and $\mu$ is a typical mass of the theory, the vacuum energy
density can be written as
\begin{equation}
<V> = V^{(0)} \ \Lambda^4 + V^{(2)} \ \Lambda^2 + V^{(4)} \ \ell n
{\Lambda \over \mu} \ .
\end{equation}
Nambu imposes the vanishing of quartic $V^{(0)}$ and quadratic $V^{(2)}$
divergences. The first condition is automatic in supersymmetric theories
and the second gives a constraint between the Yukawas known as the
Veltman condition.

In the example chosen by Nambu, for two couplings $\lambda_1$ and $\lambda_2$
, the vacuum energy reads
\begin{equation}
<V> = - A ({\lambda_1}^4 + {\lambda_2}^4) + B ({\lambda_1}^2 \ell n
{\lambda_1}^2 + {\lambda_2}^2 \ell n {\lambda_2}^2)
\end{equation}
and the Veltman condition is
\begin{equation}
{\lambda_1}^2 + {\lambda_2}^2 = a^2 \ .
\end{equation}
A picture in the case $B = 0$ is shown in Fig.1.
%%%%%%%%%%%%%%%%%%%%%%%%%%%%%%%%%%%%%%%%%%%%%%%%%%%%%%%%%%%%%
\begin{figure}
\vspace*{8 cm}
\includegraphics{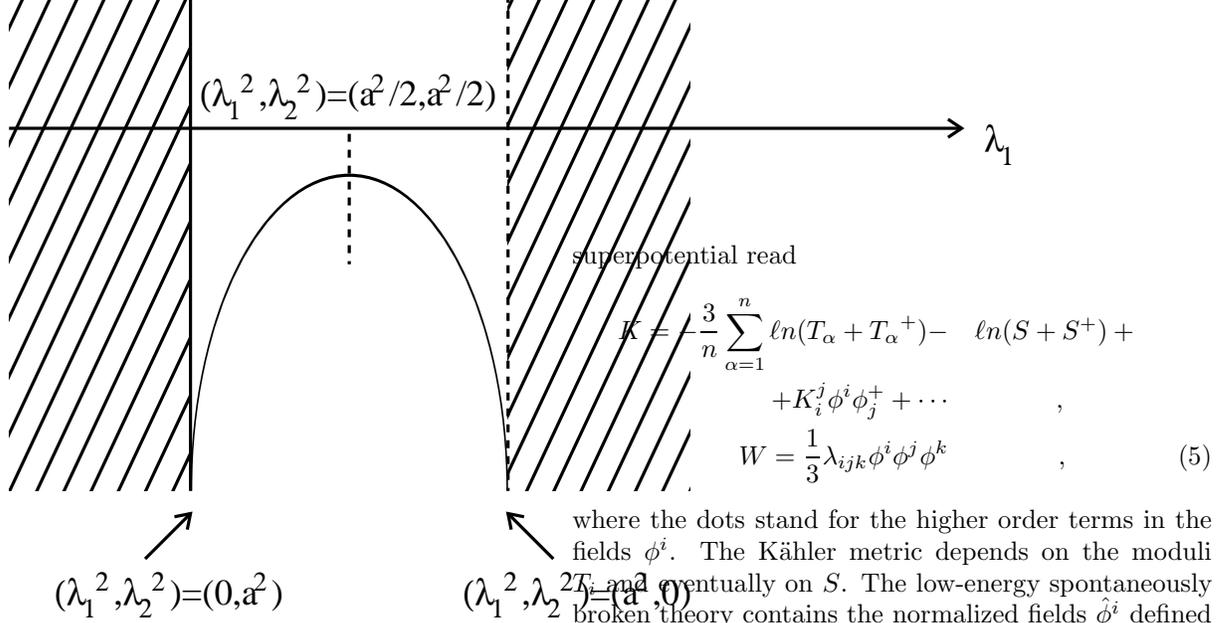}
\caption{The vaccum energy as a function of $\lambda_1$, neglecting the
logarithmic corrections.}
\end{figure}
%%%%%%%%%%%%%%%%%%%%%%%%%%%%%%%%%%%%%%%%%%%%%%%%%%%%%%%%%%%%%
The shaded regions are excluded by the Veltman condition, eq.(3). For $B
\not = 0$ new minima for $<V>$ appear close to the boundary. They
correspond to the configurations $(\lambda_1 , \lambda_2 = (a , a \ e^{-A
a^2 \over B}))$ or $(a \ e^{-A a^2 \over B} , a)$. The hierarchy is
obtained if ${A a^2 \over B} >> 1$. The mechanism is easily generalized
to more couplings, in which case {\it one} coupling $\lambda_{i_{0}}
\simeq a$ and all the others are exponentially suppressed ${\lambda_i
\over \lambda_{i_0}} << 1$, $i \not = i_0$. The applicability of this
idea to the Standard Model is under study \cite{gher}.

An analog of the Veltman condition can be obtained in superstring
models, imposing the vanishing of the quadratic divergences coming from
supergravity \cite{bd}.
%%%%%%%%%%%%%%%%%%%%%%%%%%%%%%%%%%%%%%%%%%%%%%%%%%%%%%%%%%%%%%%%%%%%
\section{Yukawas as dynamical variables in low-energy supergravity
and superstrings.}
The question of the dynamical nature of the Yukawa couplings find a
natural answer in superstrings. In these models, the Yukawa couplings
have a non-trivial dependence on the moduli fields which caracterize the
complex structure of the compact manifold. In the effective supergravity
theory, the existence of moduli manifests ussually in flat directions in
the scalar potential. They appear due to some approximate, non-compact
duality symmetries acting on the moduli. As a consequence, the vev's of
the moduli are not determined at the tree level of supergravity and the
Yukawas can be considered as dynamical variables. The duality symmetries
form an $SL(2,Z)$ group and are described by
\begin{equation}
T_{\alpha} \rightarrow {a_{\alpha} T_{\alpha} - i b_{\alpha} \over {i
c_{\alpha} T_{\alpha} + d_{\alpha}}} \ , \ a_{\alpha} d_{\alpha} -
b_{\alpha} c_{\alpha} = 1 \ ,
\end{equation}
where $a_{\alpha} \cdots d_{\alpha}$ are integer numbers and
$T_{\alpha}$ are the moduli. There are two distinct possibilities :

i) The number $n$ of undetermined $T_{\alpha} > $ The number $M$ of
Yukawas $+ 1$.
In this case we can freely perform the minimization with respect to all
the Yukawas.

ii) The number $n$ of undetermined $T_{\alpha} < $  The number $M$ of
Yukawas $+ 1$.
In this case, we will have generically geometrical constraints.

The most simple and interesting case is to have only one
constraint, corresponding to the same number of Yukawas and moduli. This
is the case which will be investigated in the next section.

Consider a model containing the dilaton-type field $S$ and the moduli
$T_{\alpha}$, specific to the superstring effective supergravities. The
K\"ahler potential and the superpotential read
\begin{eqnarray}
K= - {3 \over n} \sum_{\alpha=1}^n \ell n (T_{\alpha} + {T_{\alpha}}^+)
-& \ell n (S + S^+) + \nonumber\\
 + K^j_i \phi^i \phi^+_j + \cdots &\ , \nonumber\\
W={1 \over 3} \lambda_{ijk} \phi^i \phi^j \phi^k& \ ,
\end{eqnarray}
where the dots stand for the higher order terms in the fields $\phi^i$.
The K\"ahler metric depends on the moduli $T_i$ and eventually on $S$. The
low-energy spontaneously broken theory contains the normalized fields
$\hat \phi^i$ defined by $\phi^i = (K^{-1/2})^i_l \hat \phi^l$ and the
Yukawas $\hat \lambda_{ijk}$. In order to obtain the relation between
$\lambda_{ijk}$ and $\hat \lambda_{ijk}$, consider the scalar potential
\cite{cfgp}, which contains the piece
\begin{equation}
V \supset \ e^{K} (K^{-1})^i_j D_i W \bar D^j \bar W = \hat W_i \hat {\bar
W^i} \ ,
\end{equation}
where $D_i W = \partial W / \partial \phi^i + K_i W$ and $\hat W = {1
\over 3} \hat \lambda_{ijk} \hat \phi^i \hat \phi^j \hat \phi^k$. From
eq.(6) we get the required relation
\begin{equation}
\hat \lambda_{ijk} = e^{K \over 2} (K^{-1/2})^{i'}_i (K^{-1/2})^{j'}_j
(K^{-1/2})^{k'}_k \lambda_{i'j'k'} \ ,
\end{equation}
which mathematically express the dependence of $\hat \lambda_{ijk}$ on
the moduli through the K\"ahler potential $K$.
%%%%%%%%%%%%%%%%%%%%%%%%%%%%%%%%%%%%%%%%%%%%%%%%%%%%%%%%%%%%%%%
\section{Constraints between low energy Yukawas.}
The condition to have constraints between $\hat \lambda_{ijk}$
(combinations which do not depend on the moduli) is \cite{bid}
\begin{equation}
rang \left( \partial \hat \lambda_I \over \partial t_{\alpha} \right) <
min (M , n) \ ,
\end{equation}
where $I = 1 \cdots M$ replaces the indices $i,j,k$ and $t_{\alpha} =
T_{\alpha} + T^+_\alpha$. The condition to have just one constraint
is (in the case $M = n$)
\begin{equation}
\det \left( \partial \hat \lambda_I \over \partial t_{\alpha} \right) =
0 \ .
\end{equation}
A {\it natural} solution for eq.(9) is $\sum_\alpha t_\alpha
{\partial \hat \lambda_I \over \partial t_\alpha} = 0$, so the Yukawas
$\hat \lambda_{ijk}$ to be {\it homogeneous functions} of the moduli.
This homogeneity property translates into a scaling property for the
Kahler metric
\begin{equation}
t_{\alpha} {\partial \over \partial t_{\alpha}} K^i_j  = - K^i_j \ .
\end{equation}
This suggest us to consider the no-scale models \cite{cfkn}, which were
introduced in order to get flat directions on the scalar potential, in
connexion with the positivity of the energy in supergravity.

The class of the possible constraints is reduced if the K\"ahler space
spanned by the scalar fields is a homogeneous space, such that the
transformation of the K\"ahler potential $K$ under (4) is a K\"ahler type
transformation
\begin{equation}
K \rightarrow K + {3 \over n} \sum_{\alpha=1}^n \ell n |i c_{\alpha}
T_{\alpha} + d_{\alpha}|^2 \ .
\end{equation}
Defining $F_{\alpha} = {3 \over n} \ell n (i c_{\alpha}
T_{\alpha} + d_{\alpha}) $ and using eq.(11), we obtain the transformation
law of $\hat \lambda_{ijk}$ under (4)
\begin{equation}
{\hat \lambda_{ijk}} \rightarrow \bigl ( \prod_{\alpha=1}^n
e^{F_\alpha + F^+_\alpha \over 2} \bigr ) e^{- {n \over 6}
(F_i + F_j + F_k + h.c.)} \ \hat \lambda_{ijk} \ .
\end{equation}
As a consequence, in this case the only possible constraints are
multiplicative, in contrast with the Veltman-type constraint, eq.(3)
which is additive.

A very simple example is provided by a model containing two moduli $T_1,
T_2$, the dilaton $S$ and two observable fields $\phi^i$. The model is
defined by
\begin{eqnarray}
K= - {3 \over 2} \ell n ({t_1 - |\phi_1|^2}) - {3 \over 2} \ell n ({t_2
- |\phi_2|^2})& -\nonumber\\
\ell n (S + S^+) \ , \nonumber\\
W= {1 \over 3} \lambda_1 \phi^3_1 + {1 \over 3} \lambda_2 \phi^3_2 +
W(S) \ , &
\end{eqnarray}
where $W(S)$ is a non-perturbative contribution to the superpotential
which will fix the value of $S$ and simultaneously break supersymmetry
, as in the usual gaugino condensation scenario \cite{fgn}. The K\"ahler
potential parametrizes a $[SU(1,2) / {U(1) \times SU(2)}]^2 \times
[SU(1,1) / U(1)]$ manifold and the symmetry of the model is $U(1)^2
\times diagonal \ dilatation$. The low-energy Yukawas $\hat \lambda_i$ as
functions of the high-energy $\lambda_i$ read from eq.(7)
\begin{eqnarray}
{\hat \lambda^2_1}&= {8 \over 27} {[1 / (s + s^+)]} (t_1/t_2)^{3
\over 2} \lambda^2_1 \ , \nonumber\\
{\hat \lambda^2_2}&= {8 \over 27} {[1 / (s + s^+)]} (t_2/t_1)^{3
\over 2} \lambda^2_2
\end{eqnarray}
and the resulting constraint is obvious from eq.(14)
\begin {equation}
\hat \lambda_1 \hat \lambda_2 = {8 \over 27} {1 \over (s + s^+)} \lambda_1
\lambda_2 =fixed \ .
\end{equation}
The model is easily generalized to $n$ couplings and $n$ moduli. The
constraint (15) is valid at the Planck scale and must be run to
low-energy in order to be used in the dynamical determination of the
couplings.

To compute the vacuum energy at a low-energy scale $\mu_0 \sim M_{susy}$
we proceed in the usual way. Using boundary values for the independent
model parameters at the Planck scale $M_P$ (identified here with the
unification scale), we evolve the running parameters down to the scale
$\mu_0$ using the renormalization group (RG) equations and the effective
potential approach.
The one-loop effective potential has two pieces
\begin{equation}
V_1 (\mu_0) = V_0 (\mu_0) + {\Delta} V_1 (\mu_0) \ ,
\end{equation}
where $V_0 (\mu_0)$ is the RG improved tree level potential and
${\Delta} V_1 (\mu_0)$ summarizes the quantum corrections given by the
formula
\begin{equation}
{\Delta} {V_1} (\mu_0) = {1 \over 64 \pi^2} Str M^4 \left( \ell n {M^2 \over
\mu^2_0} - {3 \over 2} \right)
\end{equation}
In (17) $M$ is the field-dependent mass matrix containing the Yukawa
coupling dependence and all parameters are
computed at the scale $\mu_0$. The vacuum state is determined by the
equation ${\partial V_1 / \partial \phi_i} = 0$. The vacuum energy
is simply the value of the effective potential at the minimum.

A dynamical determination of the couplings and gravitino mass
$m_{3 \over 2}$ was also undertaken in \cite{kpz}. The main difference
with respect to our analysis is that in
the approach  proposed in that paper, the minimization is performed
freely, with no constraint for the couplings.
%%%%%%%%%%%%%%%%%%%%%%%%%%%%%%%%%%%%%%%%%%%%%%%%%%%%%%%%%%%%%%%%%
\section*{Acknowledgments}
I would like to thank P. Bin\'etruy for his collaboration which led to
the present work and C. Kounnas for interesting discussions.

This work was supported in part by the CEC Science project no. SC1-CT91-0729.

%%%%%%%%%%%%%%%%%%%%%%%%%%%%%%%%%%%%%%%%%%%%%%%%%%%%%%%%%%%%%%%%%%%%

%%%%%%%%%%%%%%%%%%%%%%%%%%%%%%%%%%%%%%%%%%%
%%%%%%%%%%%%%%%%%%%%%%%%%%%%%%%%%%%%%%%%%%%%


\Bibliography{8}
\bibitem{nambu} Y.\ Nambu, Proceedings of the {\it International
Conference on Fluid Mechanics and Theoretical Physics in honor of
Proffesor Pei-Yuan Chou's 90th anniversary}, Beijing, 1992; preprint EFI
92-37.
\bibitem{gher} T.\ Gherghetta and Y.\ Nambu, as cited in T.\ Gherghetta,
Proceedings of the Yukawa workshop, Gainesville, February 1994.
\bibitem{bd} P.\ Bin\'etruy and E.\ Dudas, LPTHE 94/35, hep-ph/9405429.
\bibitem{cfgp} E.\ Cremmer, S.\ Ferrara, L.\ Girardello and A.\ Van
Proeyen, Nucl. Phys. B212 (1983) 413;\\
J.\ Bagger, Nucl. Phys. B211 (1983) 302.
\bibitem{bid} P.\ Bin\'etruy and E.\ Dudas, LPTHE 94/73.
\bibitem{cfkn} E.\ Cremmer, S.\ Ferrara, C.\ Kounnas and D.\ Nanopoulos,
Phys. Lett. B133 (1983) 61;\\
N.\ Dragon, U.\ Ellwanger and M.G.\ Schmidt, Nucl. Phys. B255 (1985)
540.
\bibitem{fgn} S.\ Ferrara, L.\ Girardello and H.P.\ Nilles, Phys. Lett.
B125 (1983) 457.
\bibitem{kpz} C.\ Kounnas, I.\ Pavel and F.\ Zwirner, preprint
CERN-TH.7185/94, LPTENS-94/08, hep-ph/9406256 (june 1994).
\end{thebibliography}
\end{document}